\documentclass{emulateapj}
\shorttitle{SPITZER survey of NGC2244}
\shortauthors{Balog et al.}

\begin{document}
\title{{\it SPITZER}/IRAC-MIPS Survey of NGC2244: Protostellar Disk Survival in the Vicinity of Hot Stars}

\author{Zoltan Balog\altaffilmark{1}, James Muzerolle, G. H. Rieke, Kate Y. L. Su, Eric T. Young}
\affil{Steward Observatory, University of Arizona, 933 N. Cherry Av. Tucson, AZ, 85721}
\email{zbalog@as.arizona.edu,jamesm@as.arizona.edu, grieke@as.arizona.edu, eyoung@as.arizona.edu}

\and

\author{S. Tom Megeath}
\affil{Ritter Observatory, MS 113, Department of Physics and Astronomy, University of Toledo, Toledo OH 43606}
\email{megeath@astro1.planet.utoledo.edu}

\altaffiltext{1}{on leave from Dept of Optics and Quantum Electronics, University of Szeged, H-6720, Szeged, Hungary}

\begin{abstract}
We present the results from a survey of NGC~2244 from 3.6 to 24 $\mu$m with the Spitzer Space Telescope. The 24$\mu$m-8$\mu$m-3.6$\mu$m color composite image of the region shows that the central cavity surrounding the multiple O and B stars of NGC2244 contains a large amount of cool dust visible only at 24$\mu$m. Our survey gives a detailed look at disk survivability within the hot-star-dominated environment in this cavity. Using mid infrared two color diagrams ([3.6]-[4.5] vs [5.8]-[8.0]) we identified 337 class II and 25 class I objects out of 1084 objects detected in all four of these bands with photometric uncertainty better than 10\%. Including the 24 $\mu$m data, we found 213 class II and 20 class I sources out of 279 stars detected also at this latter band. The center of the class II density contours is in very good agreement with the center of the cluster detected in the 2MASS images. We studied the distribution of the class II sources relative to the O stars and found that the effect of high mass stars on the circumstellar disks is significant only in their immediate vicinity.
\end{abstract}


\section{Introduction}

The effect on star formation of an environment dominated by high mass stars is still an unresolved problem of astrophysics. Strong stellar winds from high mass O-, and B-type stars can trigger star formation by compressing and the interstellar material, which eventually becomes gravitationally unstable \citep{Elme77}. Also photoevaporation can heat the surface of an interstellar cloud and lead to a radiatively driven implosion, thus playing a similarly (if not more) important role in triggering star formation in this environment \citep{Adams04}. On the other hand the strong extreme ultraviolet (EUV) and far ultraviolet (FUV) radiation from the same hot stars might so vigorously photoevaporate the material around the forming stars that it truncates the star formation process. A better understanding of the photoevaporation of circumstellar material would therefore be an important step toward clarifying the overall effects of hot stars upon the process of star formation.

The discovery of "proplyds" in Orion and other high-mass star-forming regions \citep[e.g.][]{Odel93,Odel94} led to a number of theoretical studies of the photoevaporation of young circumstellar disks by external radiation fields from neighboring hot stars \citep{John98,Rich98,Rich00,Mats03,Holl04,Thro05}. These studies have shown that the outer parts of disks evaporate quickly, producing structures that match the appearance of the proplyds well. \citet{John98} presented a model for the photoevaporation of circumstellar disks by an external source of UV radiation. They applied this model to $\theta^1$ Ori C and concluded that circumstellar disks are rapidly destroyed by the external UV radiation field. Excluding viscosity effects, they predict final disk sizes of $r < 1$ AU, at distances within 0.3 pc of the illuminating source and reached about 1 million years after the onset of the UV illumination of the disk. A number of other authors \citep[e.g.][]{Rich00,Mats03,Holl04,Thro05} predict similar (or shorter) timescales but differ regarding the extent of the disk erosion. 

InfraRed Array Camera (IRAC) and  Multiband Imaging Photometer for Spitzer (MIPS) observations of high mass star forming regions are perfect tools for testing these theories. They probe the inner disk emission out to $\approx$10 AU and can cover large areas containing enough stars to allow the use of statistical methods. Here we present results from Spitzer imaging of a massive young cluster, NGC~2244
 
NGC~2244 is a young, nearby open cluster containing 7 O and 65 B stars. These high mass members of the emerging cluster are probably responsible for the evacuation of the central part of the spectacular HII region, the Rosette Nebula. The cluster has been studied at optical and X-ray wavelengths \citep[e.g.][]{Ogur81, Li02, Li03, Hetem98, Bergho02, Perez87, Perez89}. It is among the most suitable areas for studying active star formation in a high-mass environment. 

\begin{figure*}[ht]
\epsscale{.8}
\plotone{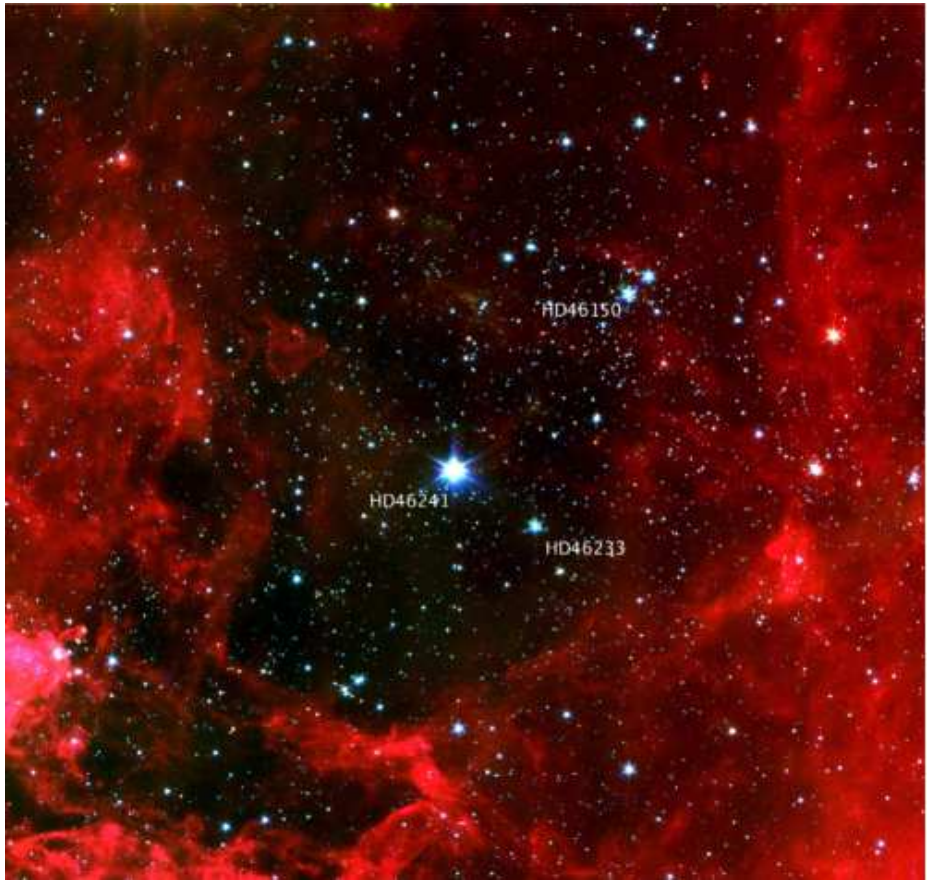}
\caption{False color image of NGC~2244. B:3.6 $\mu$m, G:4.5 $\mu$m, R:8.0 $\mu$m. The image is centered at RA=06$^{\rm h}$32$^{\rm m}$18$^{\rm s}$ and DEC=04$^{\circ}$51$\arcmin$46$\arcsec$ covering about 30'x30' area.}
\label{fig:colorIRAC}
\end{figure*}

NGC~2244 lies about 1.4-1.7 kpc from the Sun and has an age of about 2-3 Myr \citep{Park02,Ogur81,Perez87,Hens00,Perez91}. Most of the age determinations used color-magnitude diagrams in different filter-systems and with slightly different approaches. \citet{Ogur81} used the turn off point of the dereddened UBV color-magnitude digram and the position of the earliest O stars to estimate an age of $4 \pm 1$ Myr. \citet{Perez91} used the statistical method of \citet{Schr88} and derived $2.1 \pm 1.6$ Myr. Probably the most accurate age and distance estimate was performed by \citet{Hens00}. They used uvby photometric and high resolution spectroscopic observations of the cluster member and eclipsing binary V578 Mon. They applied a Fourier disentangling technique resulting in the distance and age $1.39 \pm 0.1$ kpc and $2.3 \pm 0.2$ Myr respectively.

\citet{Park02} calculated the ages of the pre-main sequence (PMS) and main-sequence cluster members (identified from the combination of photometric, spectroscopic, and proper motion studies) using the evolutionary models of \citet{Scha92} for main sequence stars and the \citet{Dant94} and \citet{Sven94} PMS models. The turnoff age of the cluster was 1.9 Myr while the mean age of the low mass PMS stars (2.5 M$_\odot$ $>$ M $>$ 0.5 M$_\odot$) was under 1 Myr with an age spread of 6 Myr, although the small numbers of PMS stars and low photometric quality for these objects makes the estimate of the age spread rather unreliable.
 
In this paper we present Spitzer IRAC and MIPS 24$\mu$m observations centered on the main OB core of NGC2244. We describe our observations and data reduction techniques in \S2. Our results and analysis are presented in \S3 and \S4, followed by our conclusions in \S5. We base our analysis on an assumed age of 3 Myr and distance of 1.5kpc.

\section{Observations}

Observations of NGC2244 were obtained on 2004 March 09 with IRAC \citep{Fazio04}. The IRAC survey covers about half a square degree centered on the main OB core. The 12s high-dynamic-range mode was used to obtain two frames in each position, one with 0.4s exposure time and one with 10.4s. The observation of each field was repeated twice with a small offset that allowed 20.8s integration time for each pixel. The frames were processed using the SSC IRAC Pipeline v13.2, and mosaics were created from the basic calibrated data (BCD) frames using a custom IDL program. 

\begin{figure*}[Ht]
\epsscale{.80}
\plotone{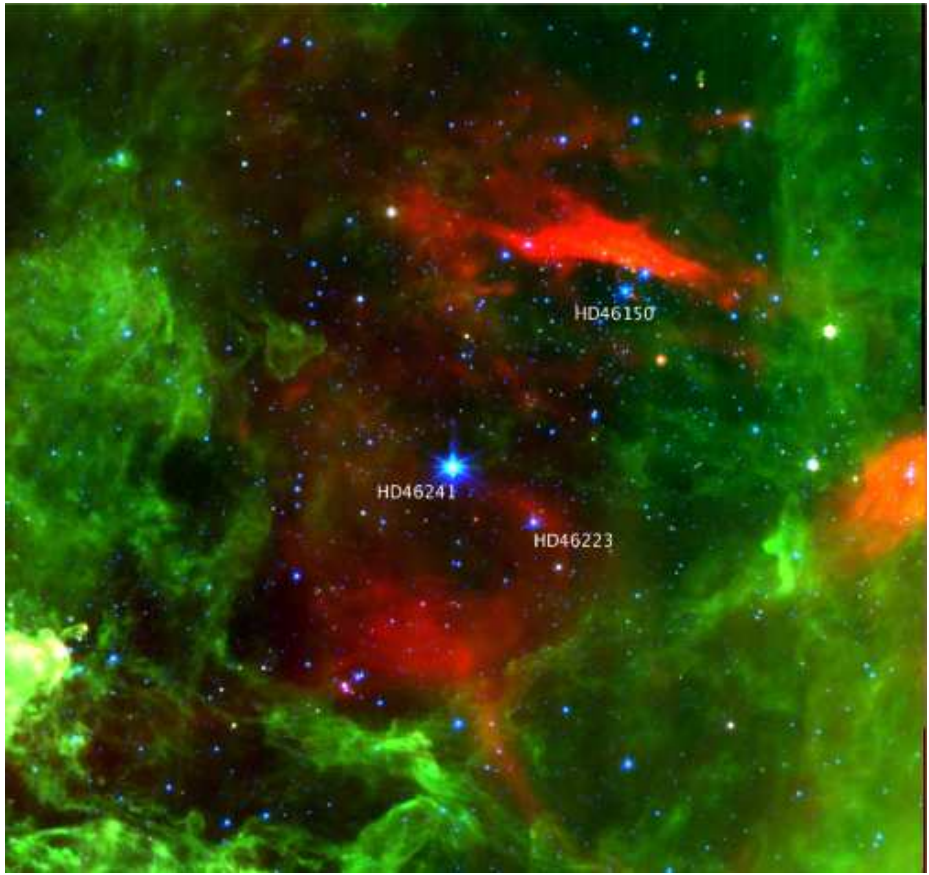}
\caption{False color image of NGC~2244. B:3.6 $\mu$m, G:8.0 $\mu$m, R:24 $\mu$m. The center and  the covered area are the same as in Fig \ref{fig:colorIRAC}.}
\label{fig:colorMIPS}
\end{figure*}

Source finding and aperture photometry were carried out using PhotVis version 1.10 which is an IDL-GUI based photometry visualization tool. See \citet{Guter04} for further details on PhotVis. The radii of the source aperture, and of the inner and outer boundaries of the sky annulus were 2.4, 2.4 and 7.2 arc-second respectively. The calibration was done by using large aperture measurements of standard star observations. The zero point magnitudes of the calibration were 19.6600, 18.9440, 16.8810, and 17.3940 for channel 1, 2, 3, and 4 respectively. Aperture corrections of 0.21, 0.23, 0.35 and 0.5 mag were applied for channels 1, 2, 3, and 4 to account for the differences between the aperture sizes used for the standard stars and for the NGC~2244 photometry.

The detected sources were examined visually in each channel to clean the sample of non-stellar objects and false detections around bright stars. We accepted as good detections those with photometric uncertainties less then 0.1, which allowed limiting magnitudes of 17.0, 16.4, 14.2 and 13.6 at 3.6 $\mu$m, 4.5 $\mu$m, 5.8 $\mu$m, 8.0 $\mu$m respectively. We detected more than 10000 sources at 3.6 $\mu$m, 8000 at 4.5 $\mu$m, and over 2000 and 1000 in the 5.8 $\mu$m and 8.0 $\mu$m images respectively. There are 1084 sources that were detected in all 4 channels, a criterion that eliminated almost all of the stars fainter than 15.0 mag at 3.6$\mu$m. We rejected the remaining eight sources fainter than 15.0 mag to help minimize the contamination from background galaxies. Based on the 2MASS magnitudes of the detected sources, the limiting mass of our survey is around 0.8-0.9 M$_{\odot}$.

The MIPS \citep{Rieke04} 24 $\mu$m survey was obtained on 2005 May 16. It covers about 0.8 square degrees, including the region of the IRAC survey. The observations were carried out in scan map mode. The frames were processed using the MIPS Data Analysis Tool \citep{Gordon05}. The high 24 $\mu$m background towards NGC2244 made automatic source extraction extremely difficult so we visually identified the sources on the background subtracted images. We identified 501 24 $\mu$m sources; most of them are concentrated in the area where the O stars are located.  PSF fitting in the IRAF/DAOPHOT package was used to obtain photometry. Our limiting magnitude at 24 $\mu$m is 10.7 using a 7.3 Jy zero-point for the 24 $\mu$m magnitude scale. 279 of the 24 $\mu$m sources were also detected in the IRAC frames with acceptable uncertainty. Almost 45\% of the 24$\mu$m sources have no counterparts within the IRAC images. There are two reasons: 1.) the 24$\mu$m survey covers a larger area than the IRAC survey (causing loss of more than 150 sources); or 2.) the photometric uncertainties of the 24$\mu$m sources are sometimes too high in the IRAC images so they are eliminated by the error cut. It is also possible that some of the 24$\mu$m sources with no counterparts are spurious detections due to the complex background, but careful visual examination of the sources limited this possibility only to a couple of cases. Many of these detections are probably high-redshift galaxies.

 Our final photometric catalogue with sources detected in all four IRAC channel sorted by right ascension is presented in Table 1.

\begin{figure*}[t]
\epsscale{0.8}
\plottwo{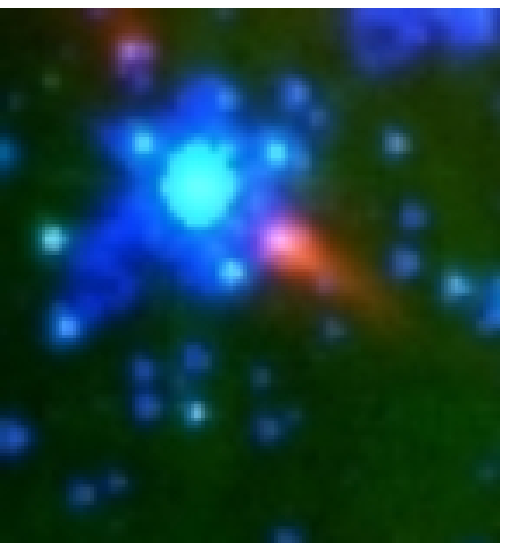}{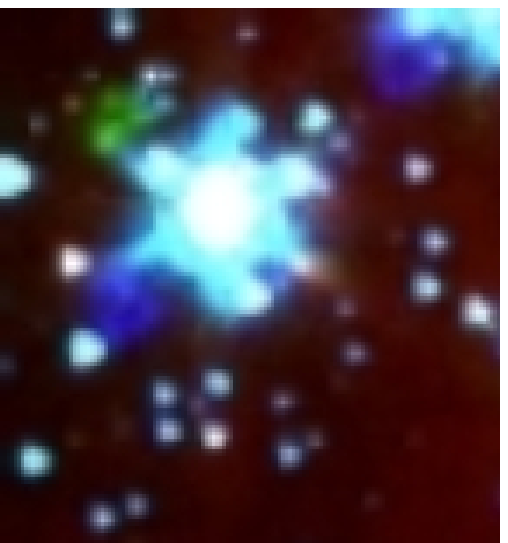}
\caption{Color images of the vicinity of HD46150 8$\mu$m. Left panel:Red 24$\mu$m; Green 8$\mu$m; Blue 3.6$\mu$m. Right panel: Red 8$\mu$m; Green 4.5$\mu$m; Blue 3.6$\mu$m. The centers of the images are at RA=06$^{\rm h}$31$^{\rm m}$55$^{\rm s}$ DEC=04$^{\circ}$56$\arcmin$22$\arcsec$, the covered area is 75$\arcsec$x62$\arcsec$.} 
\label{fig:strangesource}
\end{figure*}

\begin{deluxetable*}{llcccccccccc}[Ht]
\tablecolumns{12}
\tabletypesize{\scriptsize}
\tablewidth{0pc}
\tablecaption{The first five rows of the final photometric catalogue of sources detected in all four IRAC channels. The whole catalogue in electronic format is avalilable at the ApJ website. }
\tablehead{
\colhead{RA(J200)}&\colhead{DEC(J2000)} & \colhead{[3.6]} &\colhead{$\sigma([3.6])$} & \colhead{[4.5]} & \colhead{$\sigma(4.5])$} & \colhead{[5.8]} & \colhead{$\sigma([5.8])$} & \colhead{[8.0]} & \colhead{$\sigma([8.0])$} & \colhead{[24]} & \colhead{$\sigma([24])$}\\
\colhead{[deg]} & \colhead{[deg]} &\colhead{[mag]}&\colhead{[mag]}&\colhead{[mag]}&\colhead{[mag]}&\colhead{[mag]}&\colhead{[mag]}&\colhead{[mag]}&\colhead{[mag]}&\colhead{[mag]}&\colhead{[mag]}
}
\startdata
97.8174 &5.0398 &14.254 &0.019 &13.844 &0.020 &13.548 &0.098 &12.675 &0.092 &\nodata &\nodata \\
97.8195 &5.0555 &12.380 &0.008 &12.358 &0.009 &12.298 &0.031 &12.298 &0.050 &\nodata &\nodata \\
97.8256 &5.0197 &12.311 &0.006 &12.285 &0.008 &12.320 &0.027 &12.157 &0.053 &\nodata &\nodata \\
97.8291 &4.9182 &12.075 &0.006 &12.037 &0.008 &11.976 &0.027 &12.069 &0.082 &\nodata &\nodata \\
97.8298 &5.0017 &11.605 &0.004 &11.461 &0.005 &11.450 &0.016 &11.332 &0.038 &\nodata &\nodata \\
\enddata
\end{deluxetable*}

\section{Results}

\subsection{Extended structure}

False color images of the NGC2244 region are shown in Figure \ref{fig:colorIRAC} (B:3.6 $\mu$m, G:4.5 $\mu$m, R:8.0 $\mu$m) and  \ref{fig:colorMIPS} (B:3.6 $\mu$m, G:8.0 $\mu$m, R:24.0 $\mu$m). The prominent central cavity of NGC2244 can be identified in both images, though it is partially filled with dust emitting at 24 $\mu$m. The 24 $\mu$m emission is extremely strong in an elongated structure in the middle of the main OB core. One interesting feature of this elongated structure is that it has a bowshock-like geometry facing the nearest O and B stars. The highest mass star in the region is HD46223 (spectral type O5), which is situated southwest from the brightest star, HD46241 (K0V foreground object). There is a bow shock-like 24 $\mu$m feature associated with HD46223 but it does not seem to be as strong as the one associated with the other high mass O6 star, HD46150, and several B stars just south of the elongated 24 $\mu$m structure. These structures may be a result of stellar winds interacting with the dust.  

Another interesting feature on Figure \ref{fig:colorMIPS} is a strong 24 $\mu$m point source with an extended tail right next to HD46150 (see Figure \ref{fig:strangesource} for the enlarged images). In a separate paper \citep{Balo06}, we show that it is probably a class II source with a photoevaporating disk.

\begin{figure*}
\epsscale{0.8}
\plottwo{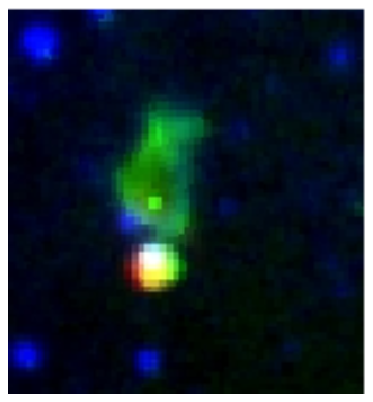}{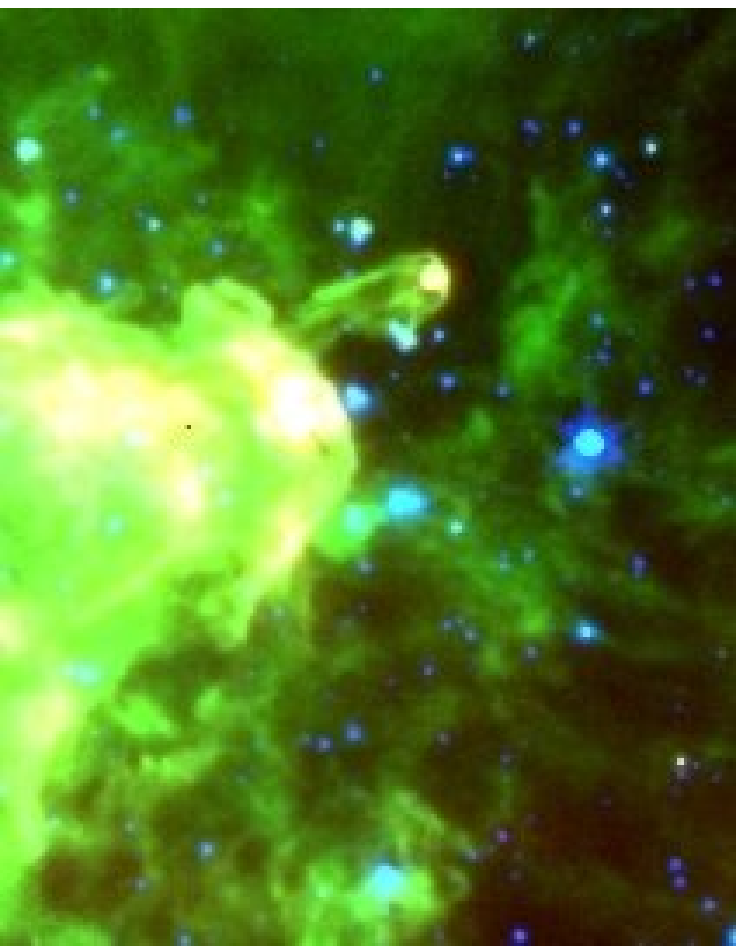}
\caption{Color images of the 8$\mu$m patch (left panel, red: 24$\mu$m; green: 8$\mu$m; blue: DSS Red, the center of the image is at RA=06$^{\rm h}$31$^{\rm m}$43$^{\rm m}$ DEC=05$^{\circ}$03$\arcmin$10$\arcsec$, the covered area is 58$\arcsec$x67$\arcsec$) and the ``pillar of creation'' (right panel, red: 24$\mu$m, green: 8$\mu$m, blue: 3.6$\mu$m,the center of the image is at RA=06$^{\rm h}$33$^{\rm m}$10$^{\rm s}$ DEC=04$^{\circ}$45$\arcmin$55$\arcsec$, the covered area is 252$\arcsec$x296$\arcsec$.}
\label{fig:globcrop}
\end{figure*}

There are two other remarkable objects in the images. One of these is a small patch on the 8 $\mu$m image extending north from a class II source (see Figure \ref{fig:globcrop} left panel for the enlarged image and Section 3.2 for the description of the classification of the sources). It seems to harbor a point source visible only at 8 $\mu$m. The structure looks like a small globule in the optical images. Very weak extended 24$\mu$m emission can be detected in association with this 8 $\mu$m feature. The second is a formation similar to a ``pillar of creation'' to the southeast from the main 24$\mu$m structure. It has a class II object at the tip, and there are a class I and several class II sources to the west of this formation (Figure \ref{fig:globcrop} right panel). This object has a linear size of 2.2 pc which is very similar to the largest of the ``pillars of creation'' in M16 \citep{Heste96}.

\subsection{Young Stellar Object Classification}

The young stellar objects (YSOs) can be characterized by their spectral energy distributions (SEDs) \citep{Adams87,Lada87}. Based on the SED we can distinguish 4 main types of YSOs: class 0, I, II, III. The class 0/I objects are deeply embedded and their SEDs peak in the submillimeter or the far infrared indicating that the source of the emission is cold dust. The class II sources are optically visible stars with infrared excess emission that is attributed to a disk surrounding the central object. Class III sources have almost no infrared excess and their photometric properties are very similar to normal main-sequence stars. We can use a mid-infrared two-color diagram to distinguish between class I, II and III objects. 

\begin{figure}[Ht]
\plotone{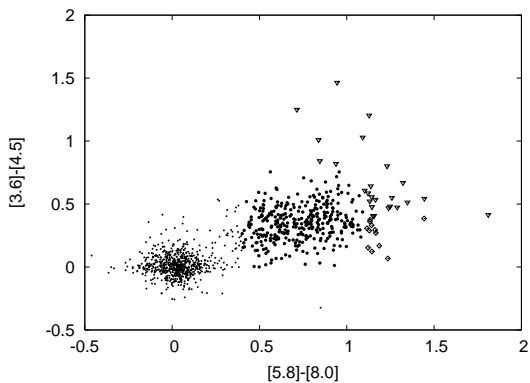}
\caption{The IRAC color-color diagram. Dots: ``normal'' stars; filled circles: class II sources; open triangles: class I sources; open diamonds: class I/II sources. The classification criteria were adopted from \citet{Allen04}}
\label{fig:IRACcolor}
\end{figure}

We identify 337 class II and 25 class I sources in the IRAC two color ([3.6]-[4.5] vs [5.8]-[8.0]) diagram using the method described by \citet{Allen04} (Figure \ref{fig:IRACcolor}). There are also 14 sources with uncertain evolutionary status ([3.6]-[4.5] $<$ 0.4 and [5.8]-[8.0] $>$ 1.1). For them, the large value of the [5.8]-[8.0] color suggests class I status while the small [3.6]-[4.5] color would indicate a class II source. We adopt the classification scheme of \citet{Megea04} and assign class I/II status to these stars. We double checked our classifications using other color-color and color-magnitude diagrams like J-H vs K-[24]  or J-H vs H-K from the 2MASS data. In each diagram the class II and class I sources occupy the regimes predicted by model calculations or by observations of other class II/I sources. 

\begin{figure}
\plotone{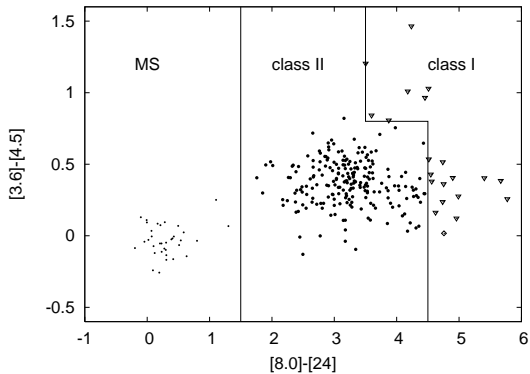}
\caption{Combined IRAC-MIPS color-color diagram. Dots: ``normal'' stars; filled circles: class II sources; open triangles: class I sources; open diamonds: outliers (stars with uncertain classification). The classification criteria were adopted from Muzerolle et al. (in preparation). Solid lines show the boundaries of each class}
\label{fig:MIPScolor}
\end{figure}

Another method of identifying stars with infrared excess (class I or class II sources) is to combine MIPS and IRAC photometry (Figure \ref{fig:MIPScolor}). This method should be more robust than the previous one because reddening hardly affects the 24 $\mu$m observations. Also, the separation of stars with different evolutionary status is larger on the combined IRAC-MIPS two color diagram. However the 24$\mu$m observations are less sensitive and so not as complete as the IRAC-only sample. Using the [3.6]-[4.5] vs [8]-[24] color-color diagram and the color criteria defined by Muzerolle et al. (in preparation) we found 20 class I and 213 class II sources out of 279 24$\mu$m sources detected in the region covered by both IRAC and MIPS observations.

We note that there are some cases where sources identified as class I on the [3.6]-[4.5] vs [8.0]-[24] plots are classified as class II sources on the IRAC two color diagram and vice versa. This indicates that the adopted boundary between class I and class II sources is uncertain and introduces some ambiguity in distinguishing the two types definitively based on their infrared colors. Where the source showed class II nature on the [3.6]-[4.5] vs [8.0]-[24] diagram but class I or class I/II on the IRAC color-color diagram, we classified it as class II because these objects are probably highly reddened class II sources. Adding these sources  to the IRAC class II sample would modify our results to 362 class II sources and 12 class I sources.

A small number of young stars might be misidentified background planetary nebulae or AGB stars or galaxies. The effects of these contaminations are usually small \citep{Megea04}. The method of Gutermuth et al. (in preparation) uses different color-color and color-magnitude diagrams to estimate the number of contaminating sources. It shows that the number of sources that might be AGN or PAH emitting galaxies is below 1\%. It is also possible that a 24$\mu$m source invisible at shorter wavelengths is projected onto a pure IRAC class II source confusing the classification methods. This configuration might appear as a class I object showing a large excess at 24$\mu$m on the [3.6]-[4.5] vs [8.0]-[24] plot while on the IRAC two color diagram it seems to be a normal class II object. The probability of such an event is extremely low (P $<$ 10$^{-6}$). Chance coincidences with background 24 $\mu$m-emitting galaxies are also possible, but unlikely in the relevant flux density range for class I sources, given typical galaxy number counts \citep{Papo04} and our flux limits.

\begin{figure*}
\epsscale{.80}
\plotone{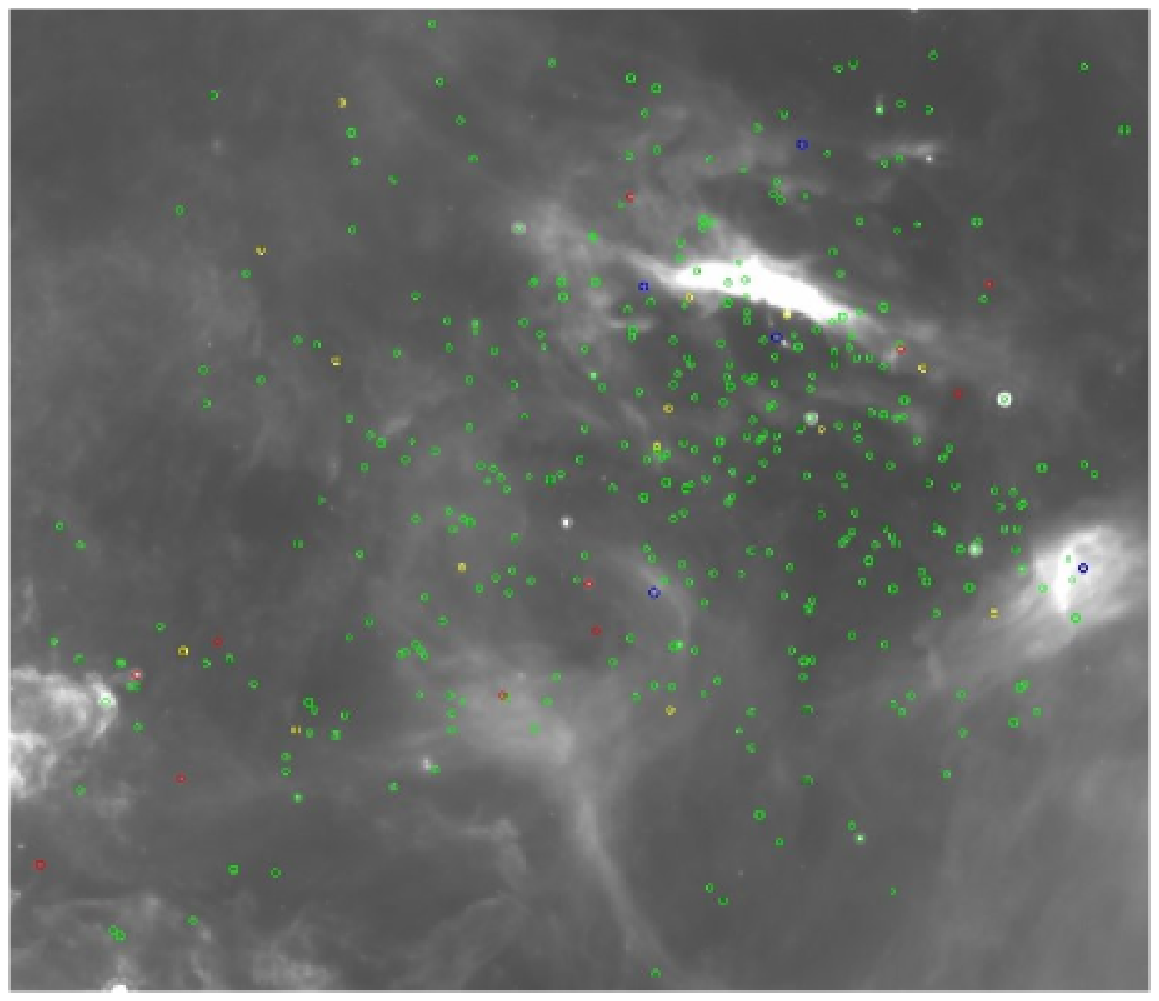}
\caption{24 $\mu$m image with stars with IRAC color excess (class I sources: red; class II sources: green; class I-II: yellow) and O stars: blue overlaid}
\label{fig:ds9}
\end{figure*}

\section{Discussion: The distribution of IR excess sources and its implications}

\begin{figure*}
\epsscale{0.80}
\plotone{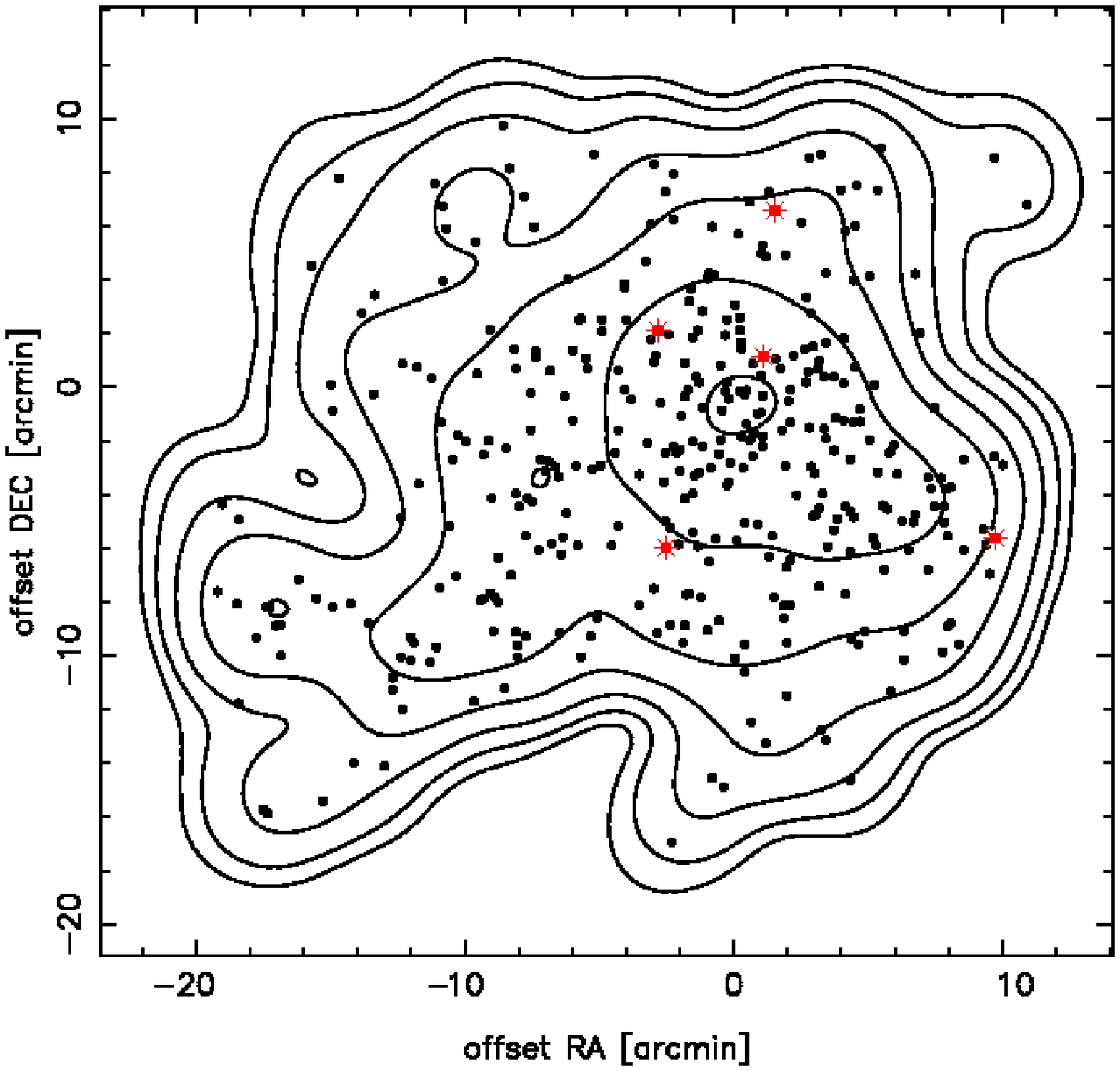}
\caption{Smoothed stellar density contours of the class II sources of NGC2244. The (0;0) coordinate corresponds to the center of the IR cluster detected in the 2MASS database by \citet{Li05}. red stars denote the known O stars detected in our survey.}
\label{fig:dens}
\end{figure*}

We show the 24 $\mu$m image with the class II and I sources and the O stars overlaid in Fig \ref{fig:ds9}. The class II sources are concentrated  mainly in the OB core. We used smoothed stellar density contours to determine the center of the class II population (Figure \ref{fig:dens}). Our calculation shows that the center of the concentration of class II sources is at RA=6$^{\rm h}$31$^{\rm m}$58$^{\rm s}$ DEC=4$^{\circ}$54$\arcmin$51$\arcsec$ (just southwest from the main 24 $\mu$m emission peak), which is in very good agreement with the coordinate of the center of the cluster derived from 2MASS data by \citet{Li05}.

\subsection{Circumstellar disks}

\begin{figure*}
\plottwo{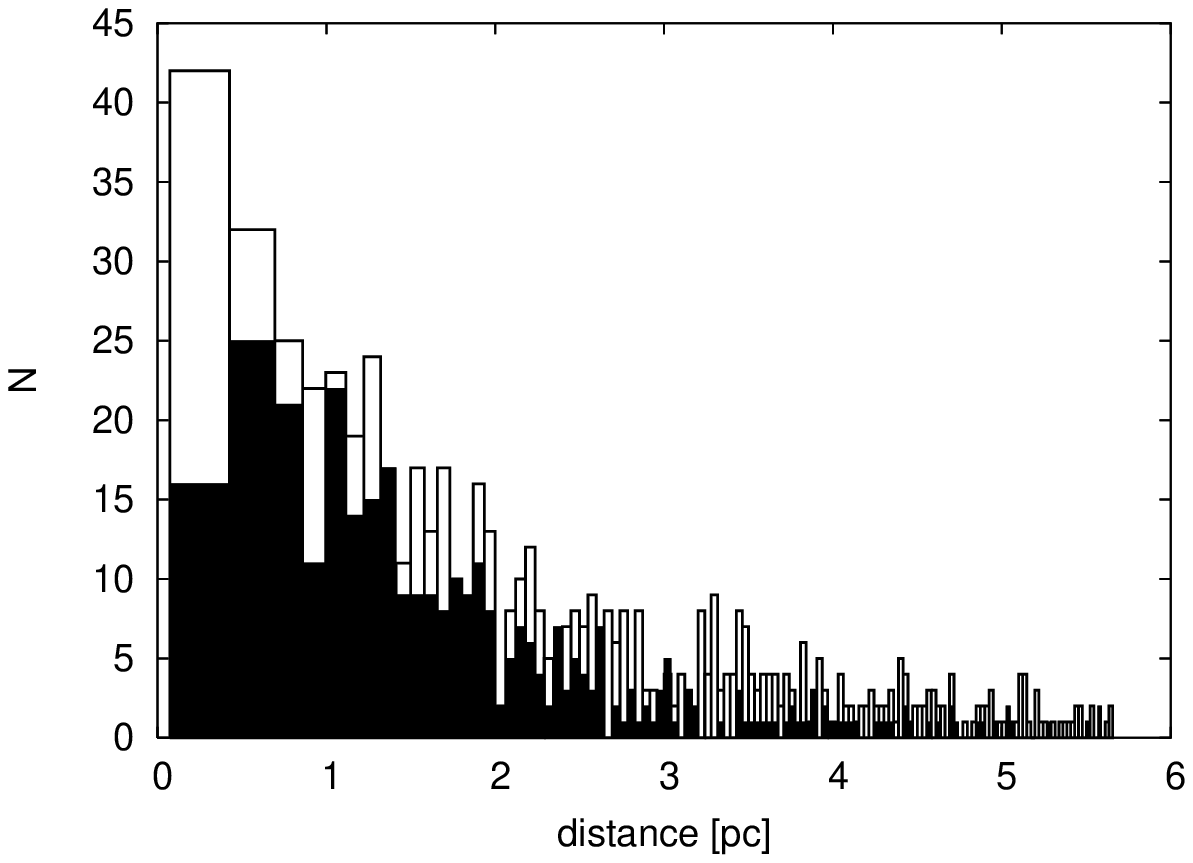}{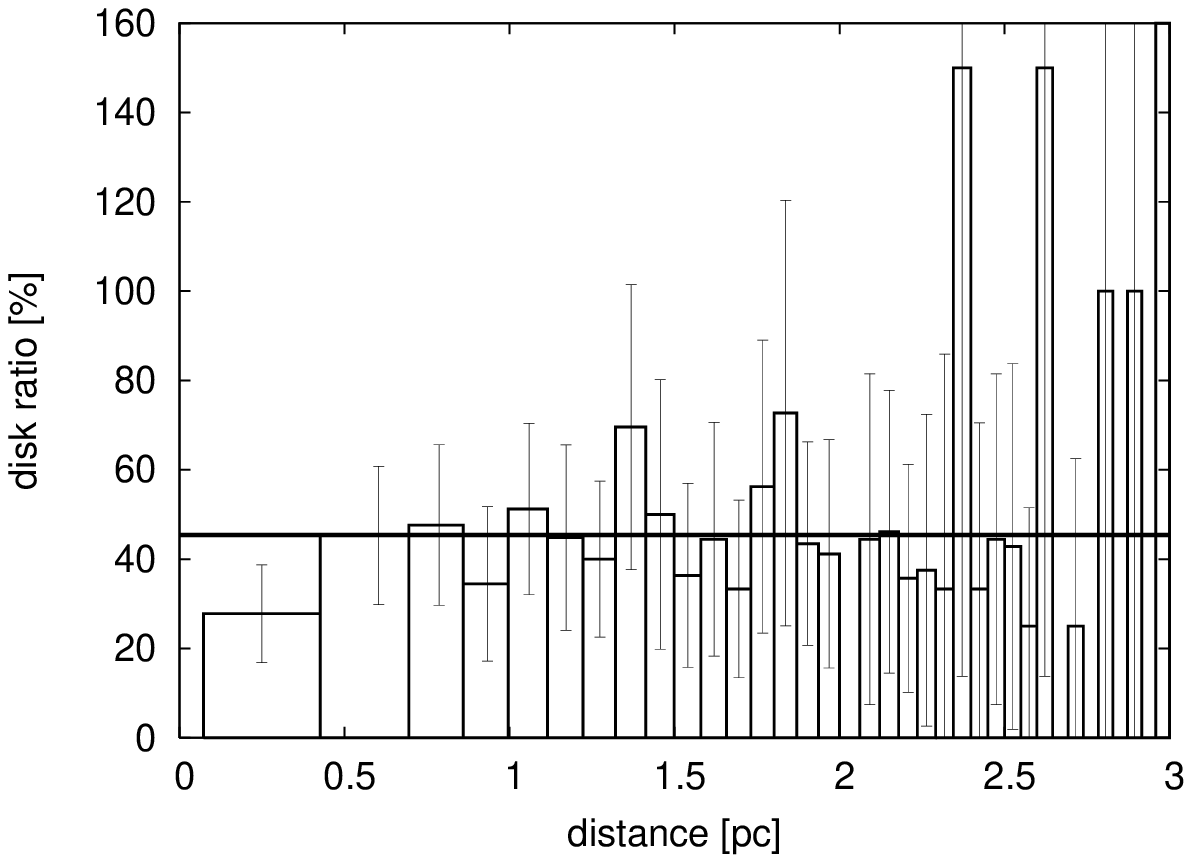}
\caption{The number of excess (filled histogram) and non excess (hollow histogram) sources (left panel) and the disk ratio (right panel) vs projected distance from the nearest O star. The solid horizontal line in the right panel shows the average disk ratio of 45.4\% between 0.5 pc and 2.5 pc. The values of more than 100\% arise when the number of non-excess sources is below the background level in some of the bins. The errorbars in the right panel represent 1 $\sigma$ values calculated through $\sigma = {\sqrt{N} \over N}$ in each bin.}
\label{fig:disthist}
\end{figure*}

The left panel of Figure \ref{fig:disthist} shows the number of stars vs their projected distance to the closest O star in equal area (0.8 $\rm pc^2$) bins. The hollow histogram represents stars without excess while the filled histogram represents class II sources. The right panel of the same figure shows the fraction of stars with excess vs their distance to the closest O star. Given that the infrared excess is generally attributed to radiation from a circumstellar disk, we consider each excess source a star with disk. There is an excess of sources without disks in the first distance bin (42 without disk vs 16 with disk) very close (d $<$ 0.5 pc) to the nearest O star. In the other bins the disk ratio seems to maintain a constant level. The number of non-disk stars as a function of projected distance from the nearest O star as shown in the left panel of Figure 9, levels off around 2.5 - 3.0 parsec. This suggests that the stars outside this radius are probably non-members and belong to the foreground or background population.

We can divide our data into three subsamples: 1.) the first bin very close to the O stars, 2.) between 0.5 and 2.5 pc from the O stars where the population of the cluster is significant; and 3.) farther than 2.5 pc from the O stars where the background and foreground population takes over. We subtracted the average background population (1 source/bin in the disk and 3 sources/bin in the non-disk population) from each bin and calculated the disk ratio in the two other regions. The disk ratio is about 27\% if the distance from an O star is less than 0.5 pc. In the remaining region where the cluster population is significant the disk ratio is about 45.4\% $\pm$ 6\%. The overall disk fraction in the cluster is 44.5\%. This is somewhat smaller but in reasonable agreement with the results of \citet{Hais01}, who found a typical disk fraction $\sim$50\% for 3 Myr-old regions using K-L excess as an indicator of circumstellar disks. \citet{Mamaj04} found the same disk fraction at 3 million years using N-band excesses. The bin to bin fluctuation is on the order of 5-10\% in agreement with the average statistical error of the disk ratio.  The small variation with radius indicates that the disk fraction is not correlated with the distance from the O stars outside a certain distance. However, close to the O stars we observe a drop of the disk fraction. To calculate the significance of this drop, we used the binomial theorem. We assumed that the sources close to and far
away from the O-stars were drawn from the same parental distribution,
that is, that the average proportion of disks was the same in both
samples and that the apparent difference was due to statistical
fluctuations distributed as in the binomial distribution. We adjusted
the common value of the average to maximize the probability, and
found that there was less than a 2
the O stars having the same intrinsic average proportion of disks
as for the far region. This behavior may be consistent with the model calculations of \citet{John98} showing that a disk can be destroyed in a couple of million years if it gets too close ($<$ 0.3pc) to an O6 star. However, there is only a trend against disks near the O stars, not a complete lack of them. Four out of five O stars in our field have at least one confirmed class II source within this radius, including the two highest mass O stars, HD46223 (O5) and HD46150 (O6). Altogether we see eight class II sources within a projected radius of 0.3 pc from one of the five O stars. Including the photoevaporating disk candidate mentioned in section 3.1, we see a net of nine class II sources in the close vicinity of an O star. 
 
We examined the possibility that these sources just seem to be close to the O stars and actually are in front of or behind them outside the 0.3 pc radius. We performed a simple Monte Carlo simulation placing 5 O stars and 362 class II sources in a sphere with a radius of 6.5 pc and counted the class II stars that were inside the 0.3 pc projected radius of any O star. After repeating the experiment 100000 times, to get accurate statistics, we found the most probable number of false companions is 3 or less (more than 50\% of all the cases). The probability of seeing 8 or more fake companions is under 3\%. We repeated the calculations using a 0.5 pc radius to reduce the effect of low number statistics and to compare our calculation with the results shown on Figure \ref{fig:disthist}. In NGC~2244 we see 16 class II sources within a 0.5 pc radius of an O star. In this case our experiment shows that the most probable number of chance alignments is 9. The  probability of seeing 16 or more stars in this region is about 4\%. 

Given the results of the simulations we can rule out the possibility that all of our class II sources are outside the radius of influence of the O star and are only chance alignments. To explain the presence of class II sources we consider the scenario that the low mass star is just passing by the O star and recently entered the ``danger zone'', making the infrared excess a short lived phenomenon. In this case, given the average velocity dispersion in the cluster ($\sim$10km/s, \citet{Liu89} or 6.8 km/s, \citet{Liu91}) and assuming radial orbits, a low mass star would spend approximately $\sim 3-4 \times 10^4$ yr in the vicinity of the O star as it passes. (The sample of \citet{Liu89} contains only 3 stars while the sample of \citet{Liu91} contains only 8 stars with sensible radial velocities, which makes the radial velocity dispersion rather uncertain.) Models by \citet{Rich98,Rich00} predict that this circumstance will expose a large amount of dust that is entrained in the evaporating gas, leading to bright IR emission. The Orion proplyds \citep{Odel93,Odel94,Smith05} are examples, and Figure \ref{fig:strangesource} shows a similar object in NGC~2244. They can be identified in MIPS images by long ``cometary'' tails of warm dust as the photon pressure ejects the grains from the vicinity of the hot stars.

Although the majority of the class II sources in the vicinity of the O and B stars do not have the expected morphology they might still be ``cometary'' sources seen head on or tail on. The modeling of the object in Figure \ref{fig:strangesource} shows that its tail lies nearly on the plane of the sky \citep{Balo06}. At the distance of NGC~2244 about 3-5 out of 6 such sources would be unresolved with MIPS given the 6'' beam size and assuming a random distribution of tail orientations. 

To sum it up, chance alignments ($\sim$9$\pm$1) and short lived unresolved cometary structures ($\sim$11$\pm$3) would account for most of the class II sources very close to the O stars (d $<$ 0.5). The drop in the portion of class II sources also suggests the presence of some objects with truncated disks.  

Outside a certain radius (0.5 pc in our case) the disk fraction does not seem to depend on the distance from the O stars. The change in disk properties near 0.5 pc may let us test models of photoevaporation. For this regime we can consider  the limiting case of the disks being long-lived. In the current photoevaporating models the behavior of the inner zones (within 50 AU of the central star) differs substantially among the studies.  For example, \citet{John98} calculated that the disks evaporate quickly down to a radius of about 20 AU, and thereafter more slowly to one of about 2 AU. They argue that effects of viscosity not included in the calculations would probably result in complete destruction of the disks. The calculations of \citet{Rich00} generally show the evaporation slowing substantially in the 50 to 100 AU region (but they only follow the evolution to 10$^5$ years). \citet{Holl04} and \citet{Thro05} calculate that the evaporation slows substantially at a radius of 5 to 10 AU, requiring 10$^5$ years in the latter case and slightly longer in the former. The models of \citet{Mats03} require a few times 10$^5$ years to evaporate to a limiting radius of about 2 AU.

Some studies \citep[see e.g.][]{John98,Rich98} derive a power law relation between the photoevaporation rate and the distance from the O star ($\dot{M_{ph}} \sim d^{-\alpha}$). The exponent of this function ($\alpha$) is in the range of about 0.7 and 1.2. These quantitative predictions are amenable to test with our observations. To estimate the effect of photoevaporation on the disk fraction and to see which slope is most probable in the case of NGC2244, we performed a Monte Carlo simulation. 

Similar to the previous case where we examined the probability of chance alignments, we scattered 5 O stars in the volume of NGC2244. Then we let a star with the disk randomly walk among the O stars with a velocity of 10 km/s. We randomly assigned an initial disk mass between 0.001 and 0.1 M$_{\odot}$ to the source based on the study of \citet{Andr05}, who found that the distribution of disk masses in a similar mass range probed by our observations is roughly equal and independent of stellar mass. We calculated the photoevaporation rate in every 0.05 pc step based on the distance from the nearest star and the photoevaporation rate - distance relation of \citet{Rich98} ($\dot{M_{ph}} = 1.46 \times 10^{-6} M_{\odot}yr^{-1}  ({ d \over {10^{17} cm}})^{-\alpha}$). They consider cases that yield $\alpha$=0.767 or $\alpha$=1.134; we therefore have let $\alpha$ vary between these values. The output of our simulation was the total mass loss and the difference between the lost mass and the initial disk mass after 3 million years. If the latter was less than zero we considered the disk totally evaporated. This very simple model does not take into account that the photoevaporation may slow down after reaching a certain radius \citep{Holl04,Thro05} and also ignores the effect of gravity (when we calculate the trajectory of the stars), the cumulative effect of more than one O star's radiation field and the central star's own UV radiation field. Other than the first, the ignored effects would speed up the photoevaporation, so our calculated disk survival rates are upper limits (unless in fact photoevaporation slows at some radius $\geq$ a few AU).  

To get reasonable statistics we ran the simulation 10000 times for the same O star configuration changing the initial mass of the disk and the initial position of the disk-bearing star. Then we repeated the experiment several times for different O star configurations. We also considered two extreme cases: 1.) the disk-bearing stars could freely leave the cluster; and 2.) the disk-bearing stars bounce back from the cluster boundaries. We found that the difference between the two cases is within the error of our simulations, which is on the order of 2-3\%. 

Given the fact that the velocity dispersions of \citet{Liu89} and \citet{Liu91} are very uncertain and much larger than the virial velocity ($\sim$ 1 km/s) of the cluster, we repeated the experiment using the velocity dispersion of 2 km/s measured in the Orion Nebula Cluster \citep{Sici05}. The effect of this velocity change on the disk survival ($\sim$1\% for each slope we studied) is smaller than our errors so we use our original values in the following. 

Assuming that a 3 million year old cluster without O stars has a disk fraction around 50\% \citep{Hais01,Mamaj04}, we found that the value of $\alpha = 0.767$ suggested by \citet{Rich98} resulted in too few surviving disk (5\%). As we make the slope steeper we see that the disk fraction initially rapidly rises (36\% with $\alpha = 1.05$), and then the rise slows down giving a disk fraction of  41\% at $\alpha = 1.134$. In spite of the weaknesses of our simple model, we can clearly conclude that the data favor the steeper slope ($\alpha > 1$) in the photoevaporation rate vs distance relation. Together with the small difference ($<$ 10\%) between the disk fraction of clusters with and without O stars and the fact that the disk fraction is constant outside a certain radius from the O stars, our lower limit $\alpha > 1$ indicates that the effect of high mass stars is significant only in their immediate vicinity ($d < 0.5$ pc).

\subsection{Class I objects}

Many of the class I sources  are located around the rims of the surrounding nebula, which might indicate self-propagating star formation triggered by the expanding shell driven by OB stars in the core of the cluster. However some class I sources appear to be in the evacuated region around the O stars. It is possible that some of these sources are only projected onto the cavity and actually they are located behind it. If we consider a configuration in which the shape of the interface between  the denser gas and the evacuated region is a half-sphere behind the cluster and the sources are evenly distributed on this interface, we find that projection can account for all of the sources apparently in the evacuated region. 

The fact that probably all class I sources are close to their birthplaces, the interface between NGC2244 and the surrounding gas, suggests that these objects are really much younger than the class II sources that are spread throughout the whole cluster. On the other hand class I sources represent a group of stars formed by triggered star formation so there might be young class II sources associated with them. Further observations are needed to study this problem in more detail.

The ratio of class I/class II sources is about 7\% in the IRAC-MIPS sample and 3\% in the pure IRAC sample. Both ratios are much lower than similar ratios ($\approx$20\%) in low-mass star forming regions such as Taurus and Ophiuchus \citep{Chen95}. This might indicate that the star formation activity in the immediate vicinity of NGC~2244 is coming to an end. On the other hand the position of the class I sources implies that star formation is still ongoing outside the evacuated region. This is in good agreement with the results of \citet{Li05} who found that sequential formation is taking place in the ambient molecular cloud.

\section{Conclusions}

We have conducted an infrared imaging survey at 3.6, 4.5, 5.8, 8.0 and 24$\mu$m of about 0.4 square degrees centered on the main OB core of NGC~2244. The limiting magnitudes of our photometry are 17.0, 16.4, 14.2, 13.6 and 10.7 at 3.6, 4.5, 5.8, 8.0 and 24$\mu$m respectively. From the analysis of the images and the photometry we conclude the following.

\begin{itemize}
\item There is a strong 24 $\mu$m feature in the middle of the main OB core of NGC~2244. It has bowshock-like geometry probably shaped by the stellar wind of the nearby O and B stars interacting with the cool dust in the neighborhood.

\item We identify 337 class II sources and 25 class I sources on the IRAC two color diagrams and 213 class II and 20 class I sources using the IRAC datasets combined with 24 $\mu$m observations. 

\item In both samples the class I/class II ratio is much lower than similar ratios in low-mass star-forming regions, implying that the star formation activity in the immediate vicinity of NGC2244 is coming to an end.

\item The center of concentration of the class II sources shows very good agreement with the coordinate of the center of the cluster derived by \citet{Li05} using star-counts from the 2MASS database.

\item Using the IRAC (and where available MIPS) data we find that roughly 44.5\% of cluster members have disks, which is somewhat smaller but in reasonably good agreement with the results of \citet{Hais01} and \citep{Mamaj04} for clusters with similar age. 

\item We found 16 class II sources in the immediate vicinity of O stars. We cannot completely explain this by line-of-sight coincidence (the most probable number of chance alignments is 9). However, it is reasonable to assume that these sources have disks that have not been close long enough to be destroyed by photoevaporation. They might be similar to the sources reported by \citet{Balo06} but unresolved because the tail is not on the plane of the sky. About 2/3 of these object (10-11 in our case) remain unresolved at 24 $\mu$m due to the large beamsize of Spitzer. 

\item We calculated  the distance of the class II sources from the nearest O star and from the corrected distribution we conclude that the fraction of stars with disks does not correlate with the distance outside 0.5 pc from the O stars. However there is a deficit of disks within the 0.5pc radius region.

\item For disks beyond 0.5pc from the O stars, we showed that the measured disk fraction favors the steeper slopes ($\alpha > 1$) in the photoevaporation rate vs O star distance relation; that is slopes like the $\alpha = 0.767$ value of \citet{Rich98} are unlikely in the case of NGC~2244. This together with the small difference ($<$ 10\%) between the disk fraction of clusters with and without O stars and the fact that the disk fraction is constant outside a certain radius from the O stars, indicates that the effect of high mass stars is significant only in their immediate vicinity (d $<$ 0.5 pc).

\item Allowing for projection, it is likely that all of the identified class I sources are located around the rims of the surrounding nebula. This indicates that they arise from self-propagating star formation triggered by the expanding shell driven by the high mass stars. It also implies that the class I objects are extremely young.
\end{itemize}
\acknowledgments

We thank the anonymous referee for comments and suggestions that greatly improved the manuscript.
This work is based on observations made with the Spitzer Space Telescope, which is operated by the Jet Propulsion Laboratory, California Institute of Technology, under NASA contract 1407. Support for this work was provided by NASA through contract 1255094, issued by JPL/Caltech. ZB also received support from Hungarian OTKA Grants TS049872, T042509 and T049082. The research made use of the Second Palomar Observatory Sky Survey (POSS-II) made by the California Institute of Technology with funds from the National Science Foundation, the National Aeronautics and Space Administration, the National Geographic Society, the Sloan Foundation, the Samuel Oschin Foundation, and the Eastman Kodak Corporation.  The Oschin Schmidt Telescope is operated by the California Institute of Technology and Palomar Observatory.

\end{document}